\documentclass[11pt]{article}
\usepackage{Blois,epsfig}
\usepackage{graphics}

\def\be{\begin{equation}}
\def\ee{\end{equation}}
\def\bea{\begin{eqnarray}}
\def\eea{\end{eqnarray}}

\begin{document}

\title{WHY THE REAL PART OF THE PROTON-PROTON FORWARD SCATTERING AMPLITUDE
SHOULD BE MEASURED AT THE LHC
\footnote{ Contribution to the proceedings of the XIth International
Conference on Elastic and
Diffractive \\
\hspace*{5mm}  Scattering, Ch\^{a}teau de Blois, France, May 15 - 20,
2005, presented by T.T. Wu.}}
\vspace*{1cm}
\author{C. BOURRELY$^{(1)}$, N.N. KHURI$^{(2)}$, Andr\'{e} MARTIN$^{(3)}$, J.
SOFFER$^{(1)}$, Tai Tsun WU$^{(3,4)}$}

\address{\vspace*{2mm}
(1) Centre de Physique Th\'eorique, UMR 6207 \footnote{UMR 6207 is Unit\'e Mixte de Recherche du CNRS and of Universit\'es
Aix-Marseille I, Aix-Marseille II \\
\hspace*{5mm} and of Universit\'e du Sud
Toulon-Var, laboratoire affili\'e \`a la FRUMAN.} , CNRS-Luminy, \\
 Case 907, F-13288 Marseille Cedex 9, France\\
(2) Physics Department, Rockfeller University, New York, N.Y.10021-6399, USA\\
(3) Theory Division, CERN, 1211 Geneva 23, Switzerland \\
(4) Gordon Mc Kay Laboratory, Harvard University, Cambridge,
MA 02138, USA}

\maketitle \abstract{For the energy of 14 TeV, to  be reached at the Large Hadron Collider (LHC), we
have had for some time accurate predictions for both the real and imaginary parts of the forward
proton-proton elastic scattering amplitude. LHC is now scheduled to start operating in two years, and
it is timely to discuss some of the important consequences of the measurements of both the total
cross-section and the ratio of the real to the imaginary part.  We stress the importance of measuring
the real part of the proton-proton forward scattering amplitude at LHC, because a deviation from
existing theoretical predictions could be a strong sign for new physics.}


\vspace*{1.0cm}

   We all know that, up to now, scattering amplitudes of PHYSICAL,
strongly interacting  particles ( {\it i.e.}~baryons and mesons ) appear to satisfy dispersion
relations, while in the particular case of quarks, since there are no asymptotic  states, this
statement would be meaningless. The most general versions of  local quantum field theory lead to
proving dispersion relations \cite{EGM69} and, more  generally,
 analyticity properties in
two variables in a rather large  domain, if one makes use of the positivity
properties of the absorptive  part of the scattering amplitude
\cite{martin66}.
Furthermore one can prove the bound for the total cross section
\begin{equation}
\sigma_{total} < \mbox{Const} (\mbox{log }{s})^2\,,
\label{equa1}
\end{equation}
where $s$ is the square of the center of mass (c.m.) energy.

The first question one faces regarding the above results is the composite nature of protons and
mesons,{\it i.e.} their quark-gluon structure. Some physicists doubted that composite particles could be
described by local fields. However, Zimmermann has proved long ago that a local field operator could be
used as an interpolating field to represent a composite particle \cite{zimer}. The asymptotic free {\it in} and {\it out}
limits of this field, could then be used to obtain S-matrix elements involving composite objects like
a proton. In the sixties, it was realized that asymptotic theory, reduction formulae and standard
analyticity properties hold also when particles, in particular composite particles, are created
by polynomials in regularized local fields or local observables, acting on the vacuum.  It was also shown in
Ref.~1, that even in these cases, the scattering amplitudes are polynomially bounded, so that
dispersion relations hold in the same way as for strictly local fields. We shall take this picture as
our starting point.

Empirically, over many years, dispersion  relations have always been consistent with the measured
data for energies reached by fixed target machines ({\it e.g.} pion nucleon scattering) or colliders (ISR
and SP\={P}S colliders).  Unfortunately the measurement of $\rho$ (see Eq.(2)) at the Tevatron has
such large errors that, no useful information can be extracted from the data.

The question of what will happen at LHC energies is completely open.  Here $\sqrt{s}$ will be 30
times higher than the highest energy for which $\rho$ has been previously accurately measured. From
the point of view of some string theorists extra dimensions that are needed for string theory, could
be larger than the compact ones proposed in the early string days. Indeed, in some recent models \cite{anton04},
these could be of a scale not far above that of LHC and could introduce observable non-local effects.
However, one should note that at present, none of these string type theories have a clear
definition of a scattering amplitude.

If $a(s,t)$ denotes the spin-independent amplitude for $pp$ (and $p\bar p$) elastic scattering, where $t$ is
the momentum transfer, we define the ratio of real to imaginary parts of the forward amplitude
\begin{equation}
\rho (s) = {\mbox{Re}~a(s, t=0) \over \mbox{Im}~a(s, t=0)} ~,
\end{equation}
the total cross section
\begin{equation}
\sigma_{total} (s) = \frac{4\pi}{s} \mbox{Im}~a(s, t=0)
\end{equation}
and the differential cross section
\begin{equation}
\frac{d\sigma (s,t)}{ dt} = \frac{\pi}{s^2}|a(s,t)|^2 \ .
\end{equation}

We recall that three of us, Bourrely, Soffer and Wu (BSW), have proposed more than twenty years ago an
impact picture approach \cite{bsw03}, based on the work of Cheng and Wu \cite{cheng70}, which
describes accurately all available $pp$ and $\bar p p$ elastic data. Several predictions have been
made and, as an illustration, we show in Fig.~1 (left), the predicted cross section for $\bar p p$, in
the Coulomb Nuclear Interference (CNI) region,
compared to the UA6 data \cite{bree89}.\\
\begin{figure}[t]
\begin{center}
 \begin{minipage}{7.0cm}
 \epsfig{figure=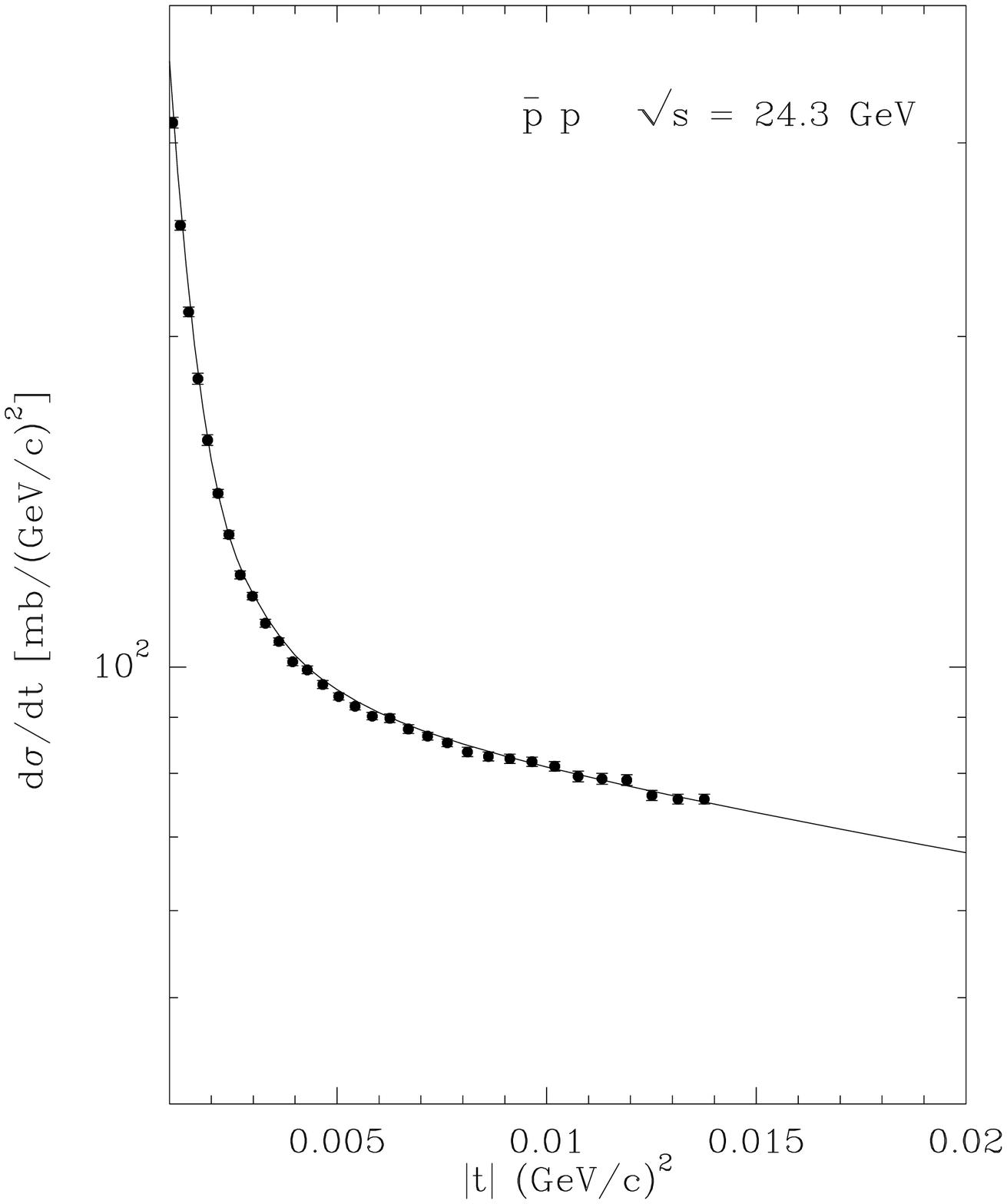,width=6.5cm}
 \end{minipage}
\begin{minipage}{7.0cm}
  \vspace*{-10.0mm}
 \epsfig{figure=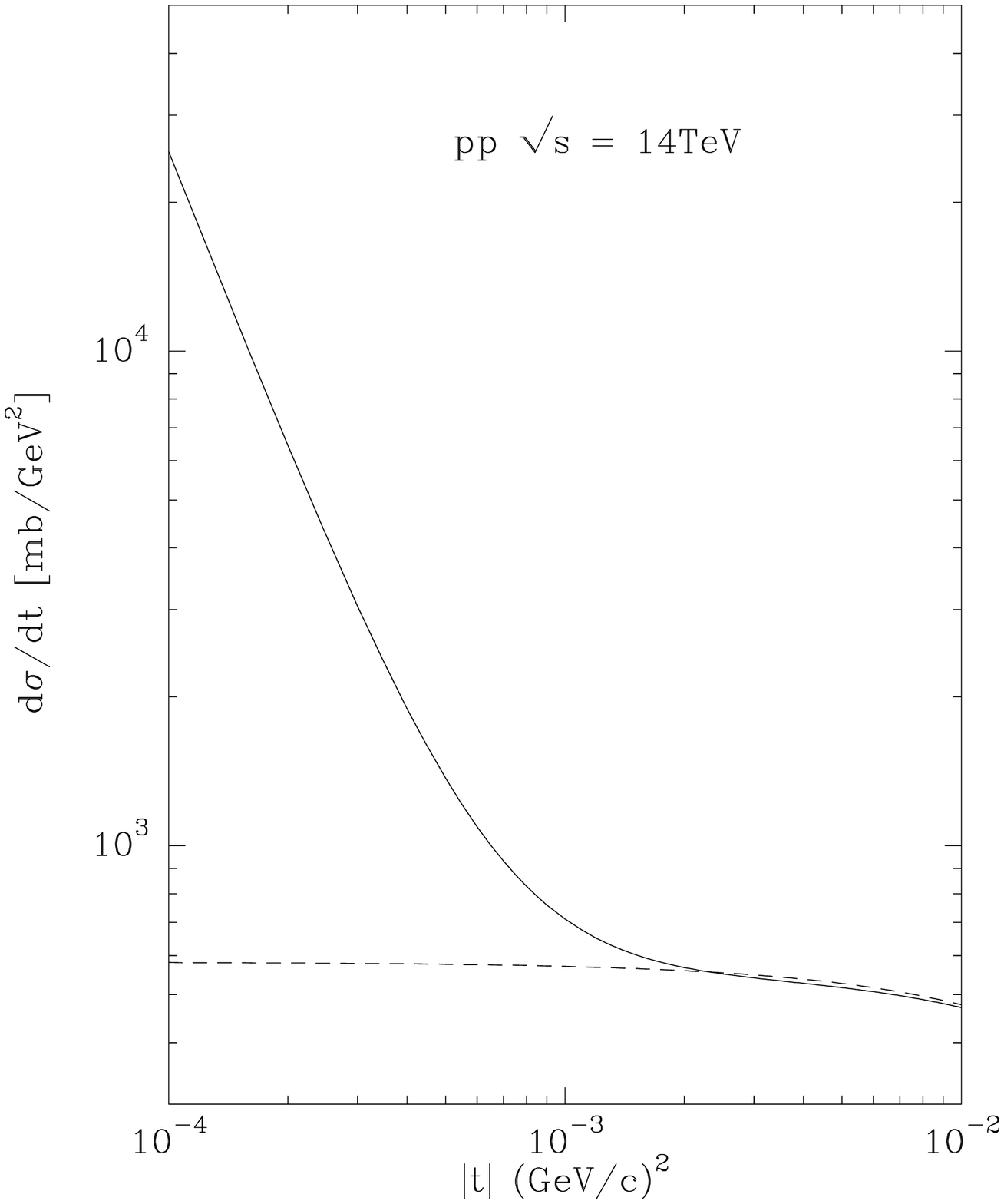,width=7.3cm}
 \end{minipage}
\begin{minipage}{14cm}
\vspace*{-9mm}
 \caption[*]{
Predictions from the BSW model. On the left, $d\sigma /dt$ for
$\bar p p$ as a function of $|t|$ in the small $t$ region for
$\sqrt{s}= 24.3 \mbox{GeV}$ compared to the UA6 data \cite{bree89}
(Taken from Ref.~[5]).
On the right, $d\sigma /dt$ for $pp$ elastic
scattering at the LHC energy, as a function of $|t|$ in the small $t$ region.
The dashed curve is the pure hadronic contribution, while the solid curve
includes both the hadronic and the Coulomb amplitudes. }
\end{minipage}
\label{fig1}
\end{center}
\end{figure}

  Let us now come back to the important question of testing dispersion
  relations which can be done in two ways :

     - use an explicit model which reproduces very well all existing data,
and satisfies, by construction, dispersion relations, such as the BSW model.

     - use fits of existing data, {\it e.g.} the one performed by the
UA4/2 Collaboration \cite{augier93}.

The superiority of the first approach is that there is no flexibility
in the predictions, while in the second it is essential to take a smooth
fit, depending on a few parameters, because otherwise, the predictive power
is lost. At the LHC energy $\sqrt{s}=14\mbox{TeV}$, the BSW model predicts
\noindent
\begin{equation}
\rho = 0.122~~~~\mbox{with}~~\sigma_{total}= 103.6 \mbox{mb}\,
\end{equation}
and for completeness we display in Fig.~1 (right), the predicted cross section
in the very small $t$ region. The UA4/2 fit predicts
\noindent
\begin{equation}
\rho =0.13\pm 0.018~~~~\mbox{with}~~\sigma_{total} = 109\pm 8 \mbox{mb}\,.
\end{equation}
We see that these numerical predictions are compatible. If the experiment gives numbers compatible
with those above, it will mean that the scale of violation is very much above the LHC energy or, that
the corresponding minimal size is much smaller. It will also mean that the predicted cross sections
of the model and of the fit are valid at much higher energies. This will allow us to have a better
idea about the magnitude of the cross sections at energies which might never be
accessible, except by cosmic ray experiments.\\

The next question is: what can one conclude if the real part of the amplitude obtained from the
dispersion integral over $\sigma_{total}$ turns out to be significantly different from the measured one?
There are three possible conclusions, all indicating new physics, that would result from such a
disagreement.  First, it could be that the total cross section beyond the LHC energy region is
radically different from what we now believe, based on the indications coming from cosmic ray data or
our expectation of a smooth slow logarithmic growth for $\sigma_{total}$.  This would be a very important
signal for new physics.  Second, it is quite possible, though less-likely, that in the gap we are
faced with, between 0.5 TeV and 14 TeV, something unexpected is happening, {\it e.g.} one or more resonances
or significant changes in $\sigma_{total}$, which is again new physics.  This "gap" was imposed on physicists by the
fear that LHC would be the last machine, and one had to go from $\sqrt{s}=2$~TeV at the Tevatron, to
$\sqrt{s}=14$~TeV at LHC. This fact makes the failure of the Tevatron $\rho$-measurement more
significant.

Thirdly, there is the possibility that dispersion relations themselves do not hold. This would be a
very significant result.

Due to the fact that no violation is seen at lower energies, the violation must be progressive and it
turns out, in our proposal as we will see, that the violation is controlled by a single parameter. We assume that
the initial analyticity domain obtained by local field theory (without the extension due to
positivity, which needs polynomial boundedness) is still valid, but that polynomial boundedness is
violated in unphysical, in particular complex, regions of the analyticity domain \cite{khuri}. It is
not so easy to implement this violation, and for instance, if one assumes a growth  like
$\exp{(\sqrt{s})}$ in complex directions, one falls back, in the end on a polynomial bound, back on
ordinary dispersion relations and back on the  standard bound on the total cross section. The first
case of non trivial violation of dispersion relations is when the scattering amplitude is allowed to
behave like $\exp{(s/s_0)}$ in unphysical and/or complex directions. If we assume that this bound
also holds inside the "Lehmann ellipse" at fixed energy, we can prove, using unitarity, that the
physical amplitudes, for $t \leq 0$, are bounded by $s^4$. This means that we can write
a dispersion integral with four subtractions. However this is not the scattering amplitude which
differs from the dispersion integral by an entire function of order one. Another way is to multiply
the scattering amplitude by a convergence factor, which guarantees that the modified amplitude has no
exponential growth in complex directions. Such a factor is
\begin{equation}
\exp{(-\sqrt{(4m^2 -s)}\sqrt{(4m^2 -u)}/s_0)} \,,
\label{equa2}
\end{equation}
a crossing symmetric term, where $u$ is the third Mandelstam variable and $m$ is the proton mass.
Because of the fact that the real part of the scattering amplitude
is indeed small in existing data, as well as in models and fits, the effects of
this exponential growth are very visible even if the scale of the
exponential, $s_0$, is much higher than the square of the LHC energy.
For instance at LHC ($\sqrt{s}$=14 TeV), as stated above,  we expect naively :
\begin{equation}
                     \rho = 0.12~~\mbox{to}~~0.13 \,.
\label{equa3}
\end{equation}

With a scale $\sqrt{s_0 }= 50 \mbox{TeV}$, the modified amplitude would lead to
\begin{equation}
\rho  = 0.21 \,.
\label{equa4}
\end{equation}

This means that we do not even need a very accurate measurement of $\rho$ to see an effect.  A
measurement with 30{\%} accuracy could be enough.  This is why we are delighted that $\rho$ will be
measured by the ATLAS detector at CERN \cite{atlas}, as a by-product of a luminosity measurement using the
Coulomb interference region.

In closing we should stress the following:  an experimental measurement of $\rho$ giving a result
consistent with Eqs. (5) and (6) will also be an important result and there will be no indication for new physics.
However, we would then have an empirical test of local field theory at length scales 30 times
smaller than what is presently known.

\section*{Acknowledgements}
We would like to thank Maurice Haguenauer and Henri Epstein for very stimulating discussions.
The work of one of us (TTW) was supported in part by the US Department 
of Energy under Grant DE-FG02-84ER40158; he is also grateful for 
hospitality at the CERN Theoretical Physics Division.\\


\vspace*{0.5cm}
\section*{References}
\begin{thebibliography}{99}

\bibitem{EGM69}  H. Epstein, V. Glaser and A. Martin,
{\it  Comm. Math. Phys.} {\bf 13}, 257 (1969).
\bibitem{martin66}  A. Martin, {\it Nuovo Cimento} {\bf 42}, 930 (1966)
and {\bf 44}, 1219 (1966).
\bibitem{zimer}  W. Zimmermann, {\it  Nuovo Cimento} {\bf 10}, 597 (1958).
\bibitem{anton04} I. Antoniadis, {\it Eur. Phys. J.} C {\bf 33}, 914 (2004).
\bibitem{bsw03} C. Bourrely, J. Soffer and T.T. Wu,~{\it Phys. Rev.} {\bf D19}, 3249 (1979), {\it Nucl. Phys.} B
{\bf 247}, 15 (1984) and
{\it Eur. Phys. J.} C {\bf 28}, 97 (2003).
\bibitem{cheng70}  H. Cheng and T.T. Wu, {\it Phys. Rev. Lett. }
{\bf 24}, 1456 (1970).
\bibitem{bree89} UA6 Collaboration, R.E. Breedon {\it et al.}, {\it  Phys.
Lett.} B {\bf 216}, 459  (1989).
\bibitem{augier93} UA4/2 Collaboration, C. Augier {\it et al.}, {\it Phys.
Lett.} B {\bf 316}, 448 (1993).
\bibitem{khuri} N. N. Khuri, Proceedings of Les Rencontre de Physique de
la Vall\'ee d'Aoste: Results and Perspectives in Particle Physics (M.
Greco, ed.), pp 701-708, {\it  Editions Fronti\`eres}, Gif-sur-Yvette, France,
1994.
\bibitem{atlas} M. Haguenauer, private communication.\\
 See also, Atlas Collaboration, I. Efthymiopoulos, these proceedings.
 \end{thebibliography}
\end{document}